\documentclass[aps,prl,floatfix,twocolumn]{revtex4}
\usepackage{graphicx,amsmath,units,xspace}

\graphicspath{{figures/}}

\newcommand{\micron}{\ensuremath{\unit{\mu m}}\xspace}
\renewcommand{\vec}[1]{\ensuremath{{\mathbf{#1}}}}
   
\newcommand{\hydrogen}{\ensuremath{\mathrm{H_2}}\xspace}
\newcommand{\oxygen}{\ensuremath{\mathrm{O_2}}\xspace}
\newcommand{\carbonate}{\ensuremath{\mathrm{HCO_3^-}}\xspace}
\newcommand{\hydroxide}{\ensuremath{\mathrm{OH^-}}\xspace}
\newcommand{\hydronium}{\ensuremath{\mathrm{H_3O^+}}\xspace}

\begin{document}                                             
                                                             
\title{Colloidal electroconvection in a thin horizontal cell: 
  I. microscopic cooperative patterns at low voltage}

\author{Yilong Han}
\affiliation{Department of Physics and Astronomy, University of Pennsylvania\\
  209 South 33rd St., Philadelphia, PA 19104}                
\author{David G. Grier}
\affiliation{Department of Physics and Center for Soft Matter Research\\
  New York University, 4 Washington Place, New York, NY 10003}

\date{\today}                                                
                                                             
\begin{abstract}
Applying an electric field to an aqueous colloidal dispersion 
establishes a complex interplay of forces among the highly mobile
simple ions, the more highly charged but less mobile colloidal spheres,
and the surrounding water.
This interplay can induce a wide variety of visually striking
dynamical instabilities, even when the applied field is constant.
This Article reports on the highly organized patterns that emerge
when electrohydrodynamic forces compete with gravity in thin 
layers of charge-stabilized colloidal spheres subjected to low voltages
between parallel plate electrodes. 
Depending on the conditions, these spheres can form into levitating
clusters with morphologies ranging from tumbling clouds, to toroidal
vortex rings, to writhing labyrinths.
\end{abstract}

\maketitle 

\section{Introduction} 
Electric fields exert forces on charge-stabilized colloidal particles
both directly through their coupling to the particles' charges and
also indirectly through their influence on the surrounding
electrolyte. Neighboring particles also interact electrostatically and
hydrodynamically with each other as they move. The resulting
cooperative motions in many-body suspensions can be quite complex, and
our understanding of such electrokinetic phenomena remains incomplete
despite more than a century of study \cite{Russel89}.

The tendency of colloidal particles to form oriented chains in
oscillatory (AC) electric fields was noted nearly a century ago
\cite{Muth27} and was correctly ascribed to field-induced dipolar 
interactions.
More recently, AC fields also have been found to organize colloidal spheres
into circulating chevron bands oblique to the field
\cite{Winslow49,Jennings90,Hu94,Isambert97,Isambert97a}, glassy bands
perpendicular to the field \cite{Larsen96}, and even highly
ordered colloidal crystals \cite{Larsen96,Larsen96a,Larsen97}.
Still other patterns form when electric fields are used to drive charged
colloid against electrodes' surfaces, with two-dimensional fractal
aggregates appeared in monodisperse suspensions
\cite{Richetti84,Wei93,Lei95,Wei95} and planar superlattices forming
in binary colloids \cite{Ristenpart03}. 
More recently, a
colloidal model system has been introduced whose phase diagram 
can be tuned with salt concentration and AC
electric field \cite{Yethiraj03}.

Even constant (DC) electric fields give rise to complex dynamics.
For example, charged spheres driven onto
electrodes' surfaces by DC fields self-organize
into epitaxial colloidal crystals
\cite{Trau96,Trau97,Bohmer96,Solomentsev97}.
DC fields also have been observed to induce complex
labyrinthine patterns
in nonaqueous nanocolloids \cite{Duan01}.

Most of these cooperative electrokinetic
effects have been explained on the basis of 
electrohydrodynamic coupling among the spheres mediated by field-induced
fluxes of ions.  
Others, such as the electrokinetic compression of 
colloidal crystals \cite{Larsen96a} have
so far resisted explanation.

In this paper, we report the remarkably diverse set of
patterns that charged colloidal particles can form as
sedimenting under gravity in a vertical DC electric field. 
Preliminary results have been published in Ref.~\cite{Han03a}. 
Beautiful, highly organized dynamic patterns with spatial periods
between 20 and 200 micrometers
form at biases near the threshold for electrolysis.
These are supplanted by macroscopic (spatial period $\sim 0.5-2~\unit{mm}$)
patterns at higher biases. 
We describe how such patterns form through the interplay of
electrohydrodynamic coupling driven by reaction-diffusion of ions
and the uniform body force provided by gravity.  We further show that
the microscopic patterns result from many-particle interactions
rather than an underlying convective instability of the electrolyte.

\section{Experimental System}

\begin{figure}[!t]
  \centering 
  \includegraphics[width=\columnwidth]{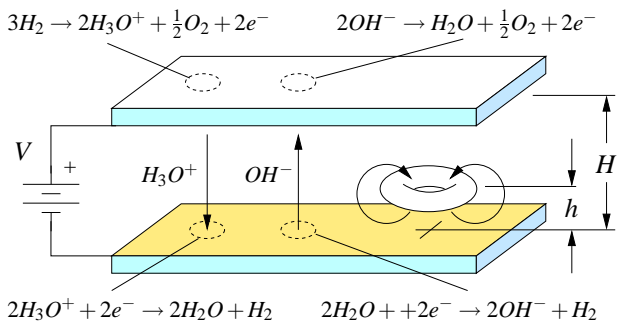}
  \caption{Parallel-plate electrochemical cell with semitransparent electrodes on
    the inner glass surfaces.  The fluxes of ions are shown schematically
    together with the electrode reactions responsible for their evolution.
    A single colloidal vortex ring is shown schematically, levitated to height
    $h$ in the inter-electrode gap $H$.  The arrows indicate the vortex's
    observed circulation direction.}
  \label{fig:schematic}
\end{figure}

The experimental apparatus is shown schematically in
Fig.~\ref{fig:schematic}. 
Our samples consist of aqueous dispersions of silica spheres
3.0~\micron in diameter (Bangs Laboratories, Lot Number 4181)
confined to slit pore between a glass microscope slide and a coverslip.
The inner glass surfaces were coated with 10~\unit{nm} thick gold
electrodes on 10~\unit{nm} thick titanium or chromium wetting layers before
assembly.  
While still optically thin, the electrodes have a
resistivity of less than 50 ohms per square and allow us to apply
uniform vertical electric fields to the confined
suspension. 
Conductive epoxy is used to make contact to the
electrodes once the cell is assembled.  
In most cases, glass spacers
were used to set the electrode separation, $H$, in the range
$100~\micron < H < 400~\micron$ across the $4 \times 1.5~\unit{cm^2}$
observation area. 
Glass access tubes bonded to holes drilled through the slides
were left open to air so that the
suspension equilibrates to a pH of roughly 5.5 through the dissolution of
carbon dioxide.
Under these conditions,
the silica spheres acquire a surface charge density of roughly
$-0.4~\unit{mC/m^2}$ \cite{Behrens01b}. 
As the density of colloidal silica is
roughly $2~\unit{g/cm^3}$, the 3.0~\micron diameter spheres
sediment rapidly onto the the lower electrode. 
Consequently, we use the areal coverage $\phi$ to
report the concentration of particles, with $\phi = 100\%$
corresponding to a close-packed monolayer and $\phi = 200\%$ corresponding to
a bilayer.
Colloidal patterns are imaged with CCD (charge-coupled device) camera
through an inverted optical microscope. 
The images and videos reported in this paper were obtained looking upward
through the lower electrode, in the direction opposite to gravity.
We define the applied DC voltage to be positive if the upper
electrode is positive so that the electric field is directed
downward, but the electrostatic forces on negatively charged particles
point upward.

The electrochemical processes in a parallel-plate water electrolysis cell
are described in most standard references,
such as Refs.~\cite{Bockris70} and \cite{Kortum65}, and
are summarized in
Fig.~\ref{fig:schematic}.
Decomposing a water molecule requires an input of 1.23~\unit{eV}, independent of
the salt concentration or the pH value of 
the solution. 
In practice, however, the redox potential is at least 0.6~\unit{V} higher
due to a voltage drop across the electrode-electrolyte interface
known as the overvoltage (overpotential). 
Once the threshold for hydrolysis is reached, the 
electrodes rapidly become coated with insulating layers of \hydrogen and \oxygen. 
These barriers soon stop the electrode reactions unless the evolved gases 
form bubbles that escape the electrodes.
A substantial nucleation barrier impedes the evolution of
\hydrogen and \oxygen bubbles
large enough to
expand and escape the electrodes. 
Consequently, a bias as large as 2.5~\unit{V} to 3.0~\unit{V}
typically is required  to maintain hydrolysis in quasi-steady-state.
The range of threshold biases is due to inevitable
variations in the surface condition of the electrodes.

Our sample cells' vacuum-deposited thin film gold electrodes yield
threshold biases between
2.6 and 3.2~\unit{V}.
Visible bubbles begin to appear between a minute and an hour after
the bias is applied, depending on the voltage.
The onset of bubble evolution
sets our observation time, and decreases with increasing voltage or salt concentration.
Although the redox potential for gold in water is only 1.5~\unit{V}, 
the overpotential prevents the anode
from etching, so that a negligible concentration of gold ions, 
$\mathrm{Au^{3+}}$, is released into solution. 
Indeed, no visible or electrochemical change in the electrodes
is seen even after several hours operation at a steady bias.
Even in the absence of added salt, however,
the electrolyte in our samples has total ionic concentrations of roughly
$n_\carbonate \approx n_\hydronium \approx 3~\unit{\mu M}$ and
$n_\hydroxide \approx 3~\unit{nM}$.
A 0.1~\micron thick Debye-H\"uckel screening layer forms at
each electrode at these concentrations, beyond which the field would not
penetrate at equilibrium.
During hydrolysis, however, the electrodes act like sources and 
sinks of \hydronium and \hydroxide ions whose currents establish
both an electric field and
a pH gradient through the cell, with the lower half volume 
becoming acidic and the upper half basic under positive bias.

\begin{figure}
  \centering 
  \includegraphics[width=\columnwidth]{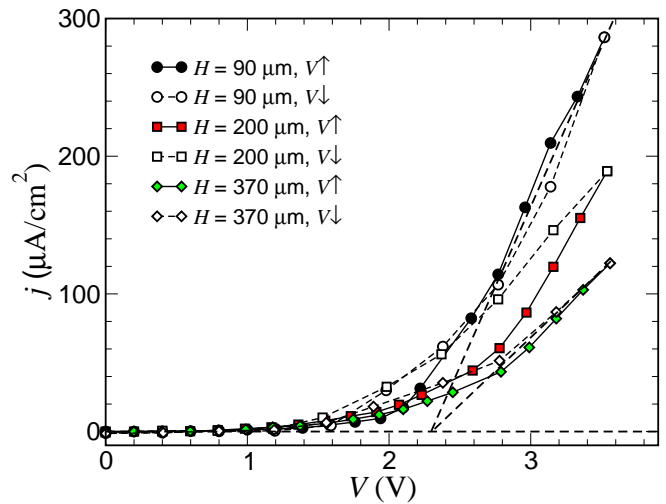}
  \caption{Current density $j(V)$ as a function of applied voltage in samples
    of deionized water in three cells of thickness $H = 90$, 200 and 370~\micron.
    A single cycle of increasing and decreasing bias is shown for each sample.
    Values were recorded after 130 sec relaxation at each voltage.
    The intersection of the extrapolated linear trends yields the
    threshold for hydrolysis.
  }
  \label{fig:VI3+-}
\end{figure}

To quantitatively characterize hydrolysis in our cell, 
we measured the current density $j(V,t)$ 
under the simplest conditions: the parallel-plate cell 
filled with deionized water without colloid.
We performed sweeping voltammetry by increasing the applied bias
in discrete steps of 0.2~\unit{V} from 0~\unit{V}
to 3.6~\unit{V} and then back down to 0~\unit{V}, 
waiting between 2 to 5~\unit{minutes} at each voltage level. 
Figure~\ref{fig:VI3+-} shows the voltage dependence of the steady-state
current in cells of different inter-electrode separations, $H$.
Each curve consists of two linear parts, one at low bias 
in which little current flows, and another ohmic regime at higher biases.
Their intersection yields 
the threshold for water decomposition \cite{Kortum65}.
Below the decomposition voltage, the small but nonvanishing current is due to the 
continuous dissolution of \hydrogen and \oxygen molecules. 
This is known as the residual current and also increases with the applied voltage. 

\begin{figure}
  \centering 
  \includegraphics[width=\columnwidth]{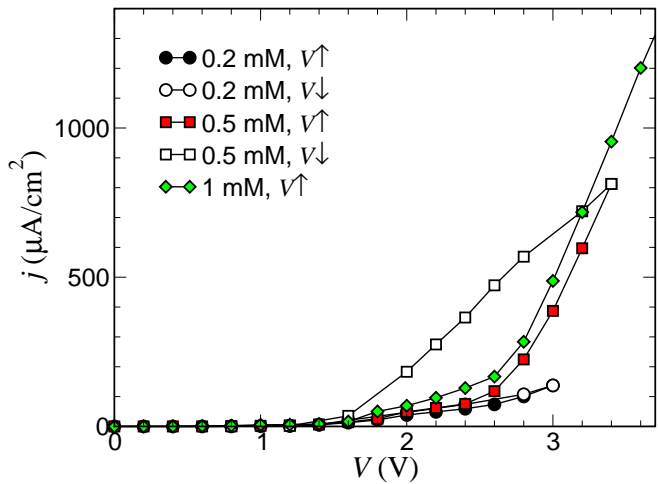}
  \caption{Current density as a function of applied voltage 
    at different salt concentrations. 
    The samples are aqueous solutions of NaCl without colloid in a cell
    $H=200~\micron$ thick.
    The current was measured after 130~\unit{sec} relaxation at each
    step in an increasing sequence of applied biases.
  }
  \label{fig:VI200salt}
\end{figure}

The higher conductivity of the high-bias regime is fed by the
additional ions from the decomposition of water.
Adding NaCl increases the conductivity but does not affect 
the threshold, as shown in Fig.~\ref{fig:VI200salt}. 
For this reason, NaCl is referred to as indifferent electrolyte.

From the slopes of the data in
Fig.~\ref{fig:VI3+-}, we estimate the equilibrium conductivity
of our electrolyte to be
$\sigma \approx 4 \times 10^{-4}~\unit{\Omega^{-1} m^{-1}}$.
Adding colloidal spheres to the deionized water 
increases the ionic concentration and thus the
conductivity.
Adding roughly $\phi = 100\%$ of silica or polystyrene sulfate (PS)
colloid increases the conductivity to
$7\times 10^{-4}$ and 
$20\times 10^{-4}~\unit{\Omega^{-1} m^{-1}}$, respectively. 
From these values, we estimated the pH 
for these systems to be 5.0, 4.7 and 4.2, respectively, with the
negatively charged colloidal spheres acting as weak
acids.

\section{Colloidal electrokinetic pattern formation}

\subsection{Interfacial crystals}
The field-driven fluxes of ions and charged particles
are coupled to flows in the surrounding electrolyte 
and can give rise to electrohydrodynamic instabilities.
The most commonly observed manifestation is 
the formation of colloidal crystals on the electrodes \cite{Trau97}.
Charged colloidal spheres are driven toward the oppositely charged
electrode, and might be expected to diffuse independently on the surface.
Instead, the particles rapidly coalesce into close-packed epitaxial
crystals \cite{Trau97}.
The strong and
long-ranged interparticle attraction 
responsible for this crystallization results from an attractive
hydrodynamic coupling induced by the 
spheres' distortion of the local fluxes of simple ions.
 
Such electrohydrodynamically bound crystals also can be induced 
by an AC field \cite{Nadal02}. 
At high enough frequencies, however, the dielectric spheres' induced
dipole moments mediate an in-plane repulsion strong enough to
and the interfacial crystals dissociate 
destabilize the interfacial crystals \cite{Nadal02}.

Field-induced surface crystals have been observed in silica, PS, 
gold and even neutral PS colloidal spheres with a wide range of diameters,
and on both indium tin oxide (ITO) and gold electrodes.
The threshold voltage for crystallization can be very low
if the spheres' diffusion is slow and if their
inherent electrostatic
and dipolar repulsions are relatively weak.

\begin{figure}[t]
  \centering 
  \includegraphics[width=\columnwidth]{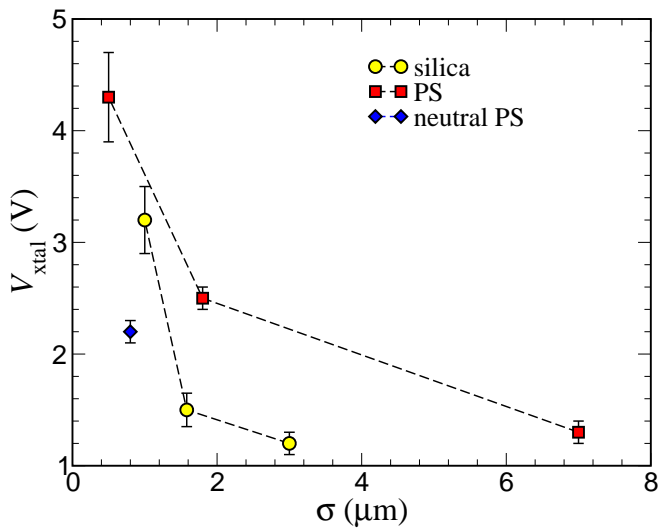}
  \caption{Threshold voltage for interfacial crystallization
    for silica, polystyrene sulfate (PS), and neutral polystyrene
    spheres over a range of particle diameters.
  }
  \label{fig:crystalV}
\end{figure}

In our system, we have observed that the threshold voltage for
field-induced crystallization
depends on the particles' mobility and charge as shown in 
Fig.~\ref{fig:crystalV}, but does not depend on the thickness 
of the cell and can be lower than the 1.23~\unit{V} redox potential of water.
Crystals were formed on the lower electrode with a negative
DC bias.
Larger spheres crystallize at lower voltages both because
they are less diffusive and also because the larger distortions
they cause in the surrounding ionic fluxes give rise to stronger
attractions.
Nominally neutral hydroxyl-terminated polystyrene
spheres reproducibly crystallize at the lowest voltages of the
particles we studied, presumably because of their substantially
weaker electrostatic repulsions.

\subsection{Colloidal vortex rings}
In this section, we examine some more complex phenomena that
arise under a more restricted set of conditions, and which
provide new insights into the interplay of 
electrohydrodynamic and other forces in macroionic systems.

Applying a positive bias in our cell tends to drive negatively charged
silica spheres upward into the bulk electrolyte. 
The field is uniform in our parallel plate geometry, so that the upward 
electrophoretic force on spheres of constant charge therefore should
be independent of height $h$ in the cell.
Under such conditions, 
spheres dense enough to sediment under gravity should remain near the 
lower electrode at low biases, 
and should rise directly to the upper electrode at higher biases 
where electrohydrodynamics forces exceed gravity.

This effect should be accentuated in charge-regulating silica spheres because
the magnitude of their surface charge increases with pH, and therefore should
increase with $h$.
Indeed, individual isolated silica spheres exhibit hysteresis in their promotion
to the upper electrode with increasing bias, and relaxation to the lower electrode
with decreasing bias, consistent with these considerations.
More concentrated dispersions, on the other hand, exhibit substantially more
interesting collective behavior.

At biases between 2.2 and 4~\unit{V}, which only slightly exceed
the threshold for hydrolysis, the background electrolyte is stable
against convective instabilities.
Nevertheless, initially quiescent collections of spheres develop
several classes of striking many-body convective patterns that highlight
a previously unsuspected role for cooperativity in charge-stabilized dispersions
subjected to electric fields.
Unlike the electrohydrodynamically stabilized surface crystals described
in previous reports, these quasi-steady-state patterns form in the bulk
of the electrolyte and reflect the interplay of electrohydrodynamic
forces and gravity.

\begin{figure}[t!]
  \centering 
  \includegraphics[width=\columnwidth]{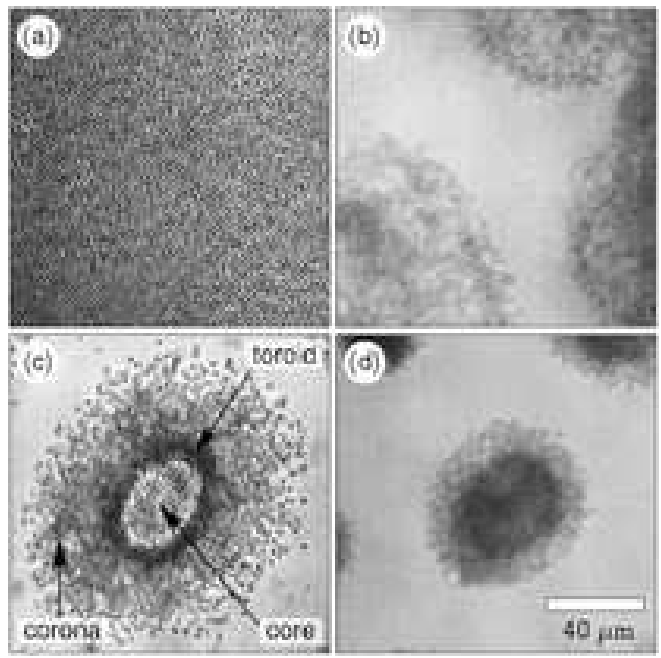}
  \caption{Typical patterns formed
    in dispersions of $3.0~\micron$ diameter silica spheres 
    at $H = 200~\micron$.
    (a) $V = 0$: Equilibrium monolayer, $\phi \approx 100\%$.
    (b) $V = 2.6~\unit{V}$: Diffuse tumbling clouds.
    (c) $V = 3.0~\unit{V}$: Circulating flower-like cluster
    levitated to $h = 40~\micron$. 
    (d) $V = 3.8~\unit{V}$: Compact circulating cluster, also
    at $h = 40~\micron$.}
  \label{fig:vortexring}
\end{figure}
Figure~\ref{fig:vortexring} shows the typical sequence of dynamic patterns 
that forms when an upward bias is
applied to dispersions of 3.0~\micron diameter silica spheres in water.
Positively biasing the upper electrode by as much as $0.8~\unit{V}$ causes
no discernible out-of-plane motion.
Abruptly applying a larger bias in the range
$0.8~\unit{V} \lesssim V \lesssim 2.5~\unit{V}$ 
causes a transient in which spheres
jump off the lower surface and then settle back to the bottom.
This transient reflects the establishment of static ionic gradients
within the water that screen out the field.

The spheres' collective behavior passes through distinct regimes as
they are driven out of equilibrium.
Just above threshold, the monolayer of spheres 
breaks into diffuse billowing clouds occupying the
lower half of the sample cell, as shown in Fig.~\ref{fig:vortexring}(b).
Pushing the system further from equilibrium might be expected to
yield increasingly chaotic behavior; but quite the opposite occurs.
Increasing the bias beyond 2.6~V coalesces the 
itinerant clouds into extraordinary flower-like clusters 
such as the example in Fig.~\ref{fig:vortexring}(c), 
all floating with their midplanes at $h = 40~\micron$ above the lower wall.
Each cluster forms around a rapidly circulating toroid whose
spheres travel downward along the inner surface
and return upward along the outside, completing one cycle in a few seconds.
The sense of this circulation is indicated schematically in 
Fig.~\ref{fig:schematic}.
Because these clusters' structure and motion 
are reminiscent of hydrodynamic vortex rings, we will refer
to them as colloidal vortex rings.

Colloidal vortex rings 
formed at higher biases tend to consist of fewer clusters,
each containing a larger number of spheres at higher density.
The dense yet vigorously circulating cluster in Fig.~\ref{fig:vortexring}(d) 
was observed at $V = 3.8~\unit{V}$.
Increasing the bias once clusters have already formed 
causes each to shrink and its circulation to accelerate.
Such over-driven clusters merge with their neighbors until they achieve the 
size appropriate to the final bias.
The clusters' height, $h$, does not change appreciably
with voltage.

By around $V = 4~\unit{V}$, some clusters' density increases enough that
they become jammed and stop circulating.
These compactified clusters still
dissociate immediately once the bias is removed.

At higher voltages, the electrolyte itself becomes 
unstable against electroconvection.
The resulting patterns span the electrochemical cell, with the colloid
acting principally as passive tracers.
We will discuss this distinct range of conditions elsewhere.

\begin{figure}[th]
  \centering
  \includegraphics[width=\columnwidth]{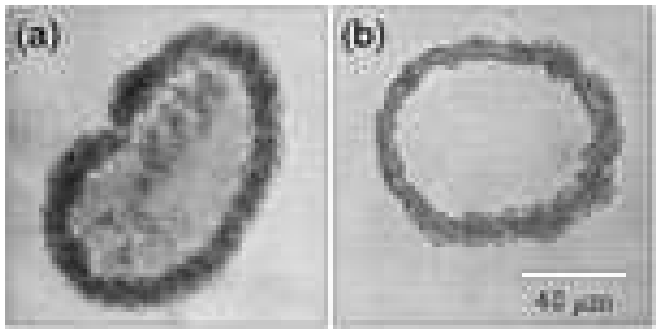}
  \caption{Bare colloidal vortex ring with and without central core.
    3.0~\micron diameter silica spheres in $H=200~\micron$ cell.
    (a) Compact toroidal cluster enclosing a domain of
    two-dimensional colloid crystal
    at $V = 2.8~\unit{V}$.  The toroid completes one cycle of circulation in 
    about 1.5~\unit{sec}.
    (b) Compact toroidal cluster with empty core.}
  \label{fig:rings}
\end{figure}

Most often, a densely packed colloidal vortex ring is surrounded by a diffuse 
circulating corona that extends outward for tens of micrometers.
Fig.~\ref{fig:rings} shows alternate forms consisting of compact
circulating toroidal clusters without coronas. 
Such bare colloidal vortex rings form most often at
slightly lower voltages, 2.6~\unit{V} versus 2.8~\unit{V},
and can coexist with fully dressed colloidal vortex rings.
These clusters also are levitated to $h \approx 40~\micron$ in
a system of thickness $H = 200~\micron$.

Although colloidal vortex rings somewhat resemble 
conventional laminar vortex rings \cite{Shariff92}, they are driven 
by quite different mechanisms \cite{Squires01} 
and also have other distinctive features.
Perhaps their most remarkable characteristic
is the presence of close-packed 
colloidal crystals at their centers.
The example in Fig.~\ref{fig:rings}(a) is particularly noteworthy because
much of its core consists of a static crystalline \emph{monolayer}.
These close-packed spheres are not flocculated, and
disperse immediately once the driving field is turned off.
Consequently, the crystals' formation and stability suggests 
the existence of a stagnation plane
along each cluster's midplane, quite unlike the streaming
flow within a purely hydrodynamic vortex ring.

The stagnation plane must end at the edge of the circulating toroid, 
and indeed particles from the core sometimes are swept 
into the circulating flow.
While the crystalline domains tend to be stable for the duration of an experimental
run (up to ten minutes at $V = 3~\unit{V}$),
some are incorporated into the surrounding toroid after a few minutes,
leaving hollow circulating rings such as that in Fig.~\ref{fig:rings}(b).

\begin{figure}
  \centering
  \includegraphics[width=\columnwidth]{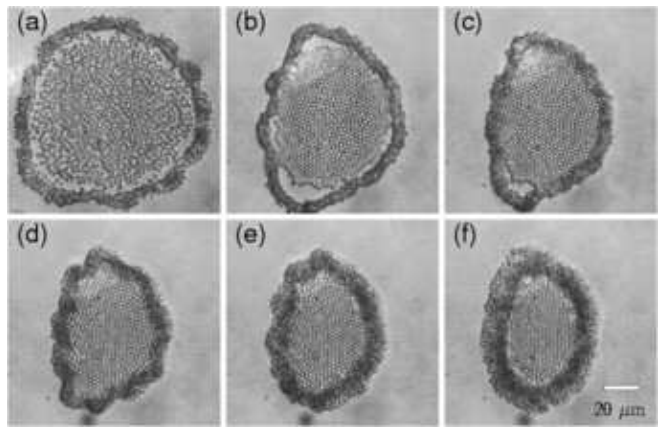}
  \caption{Central crystal formed by compression in a dispersion of
    3.0~\micron diameter silica spheres in a layer of thickness
    $H = 200~\micron$.
    The sequence of photographs show a cluster shrinking as
    the bias increases from 2.4~\unit{V} to 4.0~\unit{V}. 
    An AC component (0.7~\unit{V} square wave at 10~\unit{Hz}) 
    is applied to further stabilize the crystal. 
    The circulation period of the toroid is about 1.5~\unit{sec}. 
    (a): 0~\unit{sec}. (b): 0.5~\unit{sec}. 
    (c): 1.0~\unit{sec}. (d): 1.5~\unit{sec}.  
    (e): 2.0~\unit{sec}. (f): 3.0~\unit{sec}.  
  }
  \label{fig:shrinking}
\end{figure}

In most cases, the crystalline monolayer core evolves from a layered 
fluid-like state as shown in
Fig.~\ref{fig:shrinking}. 
As the ring's diameter decreases with increasing bias, 
the concentration of spheres in the core
increases to the freezing point, and a static crystal results. 

\begin{figure}[t!]
  \centering
  \includegraphics[width=\columnwidth]{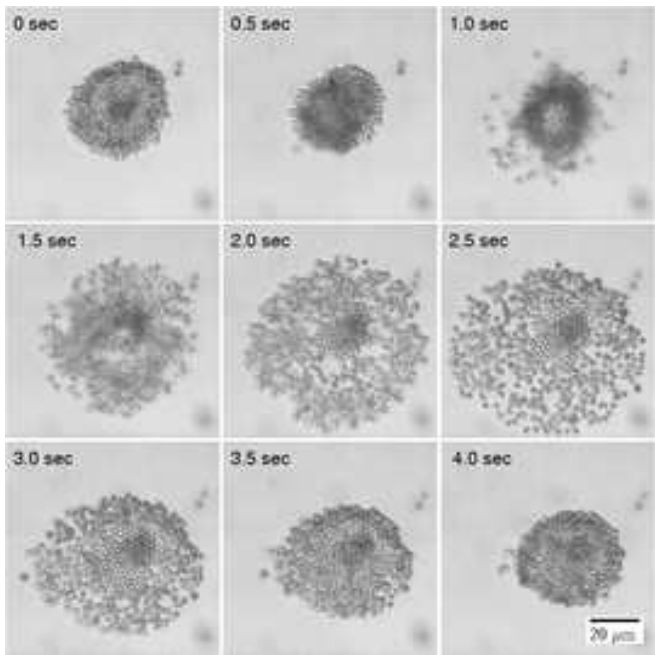}
  \caption{This sequence of images, separated by 0.5~\unit{sec}, shows one 
    period of a vortex ring's breathing mode oscillation.
    3.0~\micron silica spheres in $H = 200~\micron$ cell at 2.8~\unit{V} constant bias. 
    The cluster's core remains crystalline throughout
    the cycle.
  }
  \label{fig:oscillate}
\end{figure}

Individual colloidal vortex rings sometimes develop
breathing-mode instabilities with periods of a few seconds,
with one cycle of a typical structural oscillation appearing
in Fig.~\ref{fig:oscillate}.
The usual circulation continues throughout the oscillation.
Remarkably, the cluster's crystalline core is destroyed and reformed with
each cycle.
Neighboring clusters' oscillations do not become
phase locked, nor do they necessarily pulsate at the same frequency.
Indeed, oscillating  clusters can coexist with steadily circulating or 
compactified clusters.

\begin{figure}[t!]
  \centering 
  \includegraphics[width=\columnwidth]{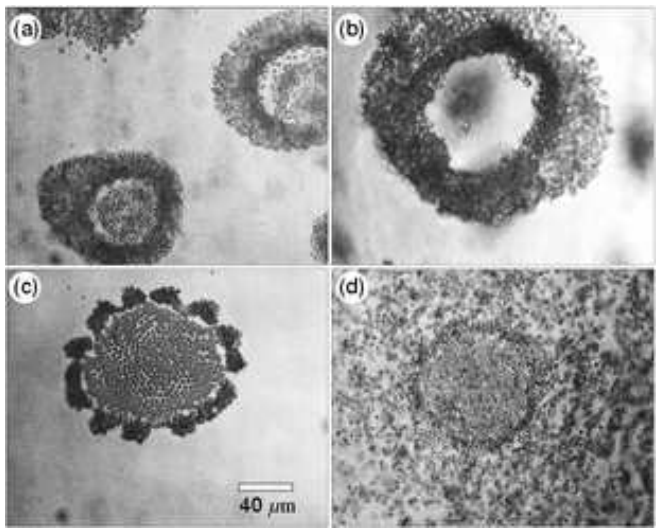}
  \caption{Variations on the theme of colloidal vortex rings. 3.0~\micron diameter silica spheres in
    $H=200~\micron$ cells.
    (a) 3.0~\unit{V}: Colloidal vortex rings with sharply defined coronas. 
    (b) 2.6~\unit{V}: Sharply bounded vortex ring without central core.
    (c) 2.6~\unit{V}: Central crystal stabilized by a 0.1~\unit{V} square wave at 1.0~\unit{Hz}
    develops scalloped boundary.
    (d) 3.0~\unit{V}: Rare variant in which the central core also is convective,
    $\phi \approx 150\%$.
  }
  \label{fig:vortexvariants}
\end{figure}

Within the range of biases for which flower-like colloidal
vortex rings reproducibly form (about 2.6 to 3.4~\unit{V}), 
we also observe a variety of variant vortex ring structures that either 
coexist with or even replace the morphologies represented in Figs.~\ref{fig:vortexring}
and \ref{fig:rings}.
Some of these are shown in Fig.~\ref{fig:vortexvariants}.

Figure~\ref{fig:vortexvariants}(a) shows
colloidal vortex rings whose coronas have sharply defined boundaries,
rather than the diffuse coronas evident in Fig.~\ref{fig:vortexring}(c).
These clusters also sometimes form without a crystalline core as can be seen in
Fig.~\ref{fig:vortexvariants}(b), and the two types can coexist.
Unlike the diffusely bounded clusters that undergo breathing mode instabilities
the corona in the diffuse morphology can undulate like a jellyfish.
Figure~\ref{fig:vortexvariants}(c) shows an extreme example of this
circumferential instability.

In all of these cases, the vigorously circulating vortex ring appears to be
virtually decoupled from the central crystalline core.
Figure~\ref{fig:vortexvariants}(d) shows a relatively rare case in which some of the particles 
in the central core are swept into the vortex ring's circulation.

In roughly half of our 
experiments, we observed vortex rings coexisting with
two-dimensional surface crystals \cite{Trau96,Trau97} that had nucleated
on the upper electrode. 
The spheres in these crystals are driven to the upper electrode by the transient
ionic current burst when the bias is initially applied.
Particles that reach the upper electrode during these transients tend to stay
there and subsequently form crystalline clusters through mechanisms that
have been previously reported \cite{Trau96,Trau97}.
Indeed, such surface crystals begin to form at biases below the onset for
vortex ring formation.
Above 2.6~V, clusters hovering in the bulk coexist with crystals trapped on the 
upper surface.
Decreasing the bias to 1.4~V, which is the threshold for forming crystals on the upper 
electrode, causes all particles in crystals to sink to the bottom.


\subsection{Worm-like colloidal vortices}

\begin{figure}[htbp]
  \centering
  \includegraphics[width=0.6\columnwidth]{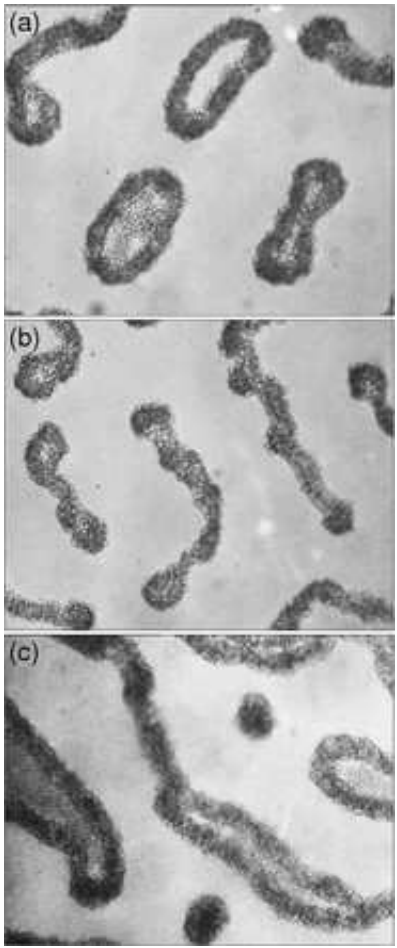}
  \caption{Worm-like colloidal vortices.
  (a) 3.4~\unit{V}: elliptical colloidal vortex rings, both with and without crystalline cores,
  coexisting with worm-like colloidal vortices.
  (b) 3.4~\unit{V}: worm-like vortices extending from vortex rings.
  (c) 2.6~\unit{V}: labyrinth of worm-like vortices.}
  \label{fig:wormlike}
\end{figure}
Not all colloidal vortices adopt a ring-like configuration, as the
examples in Fig.~\ref{fig:wormlike} show.
Lower concentration suspensions tend to favor elongated vortex rings, such as
those in Fig.~\ref{fig:wormlike}(a), that can extend into
linear worm-like structures such as those in Fig.~\ref{fig:wormlike}(b).
These clusters continue to circulate, but as linear rolls rather than vortex rings.
As a result, worm-like colloidal vortices interact more strongly that colloidal
vortex rings, and can
organize themselves into large-scale 
labyrinthine patterns, as in Fig.~\ref{fig:wormlike}(c).
As for the other patterns we have described, worm-like colloidal vortices
are levitated to roughly $h = 50~\micron$ in an $H = 200~\micron$ cell.

\subsection{Sedimented interfacial patterns}
\label{sec:microsurface}

\begin{figure}[htbp]
  \centering
  \includegraphics[width=\columnwidth]{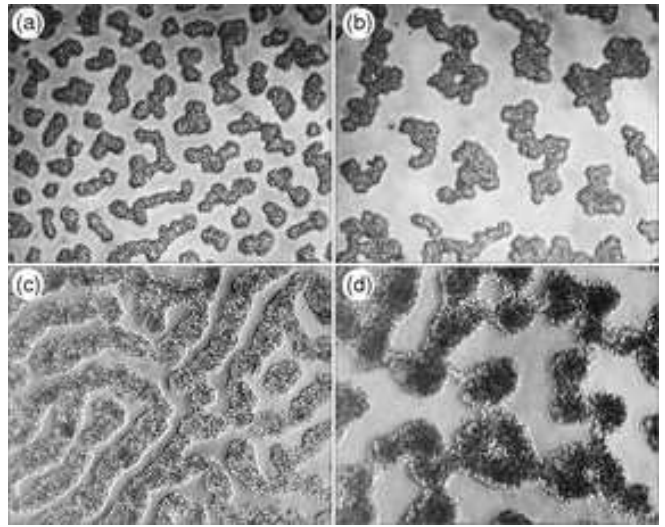}
  \caption{Sedimented interfacial patterns.
    (a) Islands of 3.0~\micron diameter silica spheres in $H = 90~\micron$ cell
    at 2.6~\unit{V}.
    (b) The same sample, after increasing the bias to 2.8~\unit{V}.
    (c) Labyrinths of 3.0~\micron diameter silica spheres in $H = 200~\micron$
    cell at 2.6~\unit{V}.
    (d) The same sample, after increasing the bias to 3~\unit{V}.
}
  \label{fig:interfacial}
\end{figure}


Although the patterns described in the previous Section form reproducibly, they occasionally
are replaced by qualitatively distinct labyrinthine and island-like patterns such as
those shown in Fig.~\ref{fig:interfacial}.
Unlike the vigorously circulating free-floating colloidal vortices,
these slow-moving clusters form near the bottom
electrode under the same conditions.
These patterns also are distinct from the previously reported interfacial
crystals, which form on the upper electrode under positive bias.

Once sedimented interfacial clusters nucleate, the choice between forming
islands, as in Fig.~\ref{fig:interfacial}(a) and 
\ref{fig:interfacial}(b), or labyrinths,
such as Fig.~\ref{fig:interfacial}(c) and \ref{fig:interfacial}(d),
depends principally on the
concentration of particles.
Two-dimensional islands form at 
surface coverages of $10\% \lesssim \phi \lesssim 60\%$, and two-dimensional
labyrinths at $60\% \lesssim \phi \lesssim 80\%$.
Three-dimensional labyrinths form at
higher concentrations. 
Islands and labyrinths can coexist at intermediate sphere concentrations.

The spheres in islands and labyrinths usually flow fluidly as the 
patterns form, in a manner that
resembles the circulation in bulk colloidal vortices.
Unlike bulk colloidal vortices, whose circulation retains its sense over time, 
labyrinthine domains sometimes flip circulation direction as they coarsen.
Unlike Rayleigh-B\'enard convection or bulk electroconvection, moreover,
neighboring domains' circulation directions generally are not correlated. 
Larger labyrinthine domains, however, develop double-roll structures,
with particles rising along the edges and sinking 
along the center line.

After a few minutes of circulation, islands and labyrinths often become jammed. 
Increasing the bias can fluidize them again, and coarsen their features.
This can be seen in the transitions
from Fig.~\ref{fig:interfacial}(a) to Fig.~\ref{fig:interfacial}(b) and from 
Fig.~\ref{fig:interfacial}(c) to Fig.~\ref{fig:interfacial}(d).

\begin{figure}[htbp]
  \centering
  \includegraphics[width=\columnwidth]{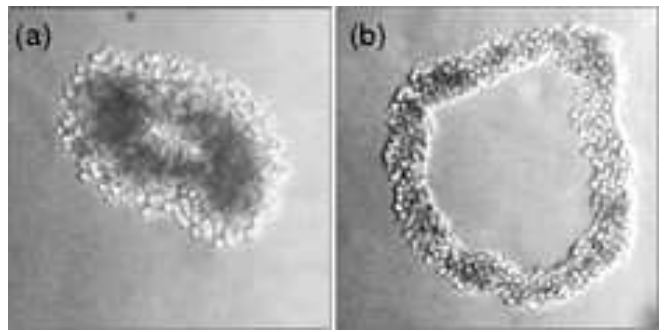}
  \caption{Sedimented interfacial vortex opening out into a smoke ring cluster.
  3.0~\micron spheres in an $H = 200~\micron$ cell.
  (a) A typical vortex ring at 3.4~\unit{V}.
  (b) The same cluster at 2.8~\unit{V}.}
  \label{fig:smokering}
\end{figure}
Increasing the bias still further causes coarsened blobs such as 
those in Fig.~\ref{fig:interfacial}(d) to circulate faster 
and develop into interfacial toroidal vortex rings, 
an example of which appears in 
Fig.~\ref{fig:smokering}(a). 
These surface-hugging clusters differ from
the free-floating vortex rings of the previous section in that they respond
irreversibly to changes in bias.
Reducing the bias on circulating blobs causes them to open out into structures
resembling smoke rings, Fig.~\ref{fig:smokering}(b).
This morphology can be formed only by first increasing and then decreasing the bias, and
clearly inherits its topology from the progenitor toroidal cluster.

The islands and labyrinths we have observed 
somewhat resemble the interfacial crystals described in 
Refs.~\cite{Trau96} and \cite{Trau97}, even though the electric field is reversed.
We conjecture that gravity maintains the particles on the lower electrode,
while the in-plane electrohydrodynamic attraction discovered in 
Refs.~\cite{Trau96} and \cite{Trau97} draws
them together into islands (Fig.~\ref{fig:interfacial}(a)) and labyrinths 
(Fig.~\ref{fig:interfacial}(c)).
Once formed, such close-packed clusters would have substantially reduced
hydrodynamic drag coefficients \cite{Happel91},
and so would remain sedimented at biases that would levitate a single sphere.
Applying the bias rapidly appears to favor the formation of levitated
clusters, presumably by disrupting this stabilizing mechanism, while slowly increasing 
the bias improves the chances to form interfacial clusters.

We have observed free-floating colloidal vortex rings
coexisting with interfacial labyrinths and islands.
This demonstrates that the two types of structures can be generated 
under the same conditions, and therefore are not distinguished by variations
in the experimental procedure.

%
%

\subsection{Anomalous dynamics in relaxing colloidal vortices}
\begin{figure}[t!]
  \centering 
  \includegraphics[width=\columnwidth]{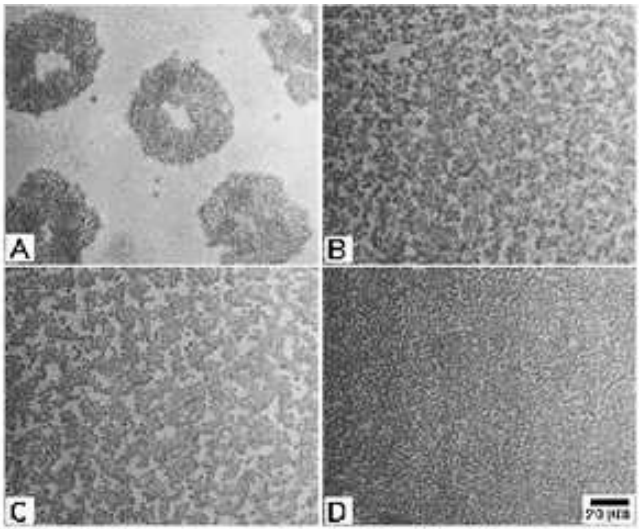}
  \caption{
    Relaxation behavior of toroidal vortex rings of 3~\micron silica spheres
    after a 4~\unit{V} bias is removed.
    (a) 0~\unit{V}: particles sedimented onto the lower electrode retain the
    basic profile of their cluster's toroidal geometry.  Left in this state,
    these particles would require several minutes to relax to their normal
    diffusive state.
    (b) Applying a positive bias of 0.8~\unit{V} levitates the
    particles slightly and allows them to relax to a uniform density.
    (c) Returning to 0~\unit{V} eliminates any diffusive motion and causes
    the spheres to form sedimented crystalline domains.
    (d) After 3 minutes, the spheres begin to diffuse and relax to a
    homogeneously random state.}
  \label{fig:m4}
\end{figure}

Turning off the driving bias after a pattern has formed frees the spheres to revert
to their equilibrium field-free behavior.
In most cases, particles sediment into a two-dimensional Brownian fluid
on the lower electrode.
On other occasions, and particularly after prolonged runs at the upper range
of biases, the sedimented particles no longer diffuse freely on the lower electrode, 
but rather sit motionlessly where they land.
This is the case for the particles in Fig.~\ref{fig:m4}(a), which had been
organized into a colloidal vortex ring before the field was switched off.
These particles reached the lower electrode rapidly enough that their toroidal
configuration is still clearly visible, and remained unchanged for several minutes.
After this period, however, the particles began to diffuse normally, suggesting
that they had not deposited onto the lower electrode.
This conclusion is supported by the observation that
a small positive bias suffices to levitate the motionless
spheres, as shown in Fig.~\ref{fig:m4}(b).
Once the positive bias removed, however, the particles sediment again
to the lower electrode and
form small crystals as they settle, a typical result appearing in Fig.~\ref{fig:m4}(c).
The spheres in these self-organized crystals still
display no Brownian motion for several minutes, after which the revert to their
equilibrium behavior, as shown in
Fig.~\ref{fig:m4}(d).
This bizarre transient behavior appears not to have been described before,
and its origin is unknown.


\subsection{Dependence on control parameters}

Similar patterns and trends emerge if the bias voltage is applied
abruptly or increased gradually enough that diffuse tumbling clouds
form before coalescing into stable clusters.
The typical number of particles in a steady-state cluster tends to
increase with the bias. 
Slightly changing the bias once clusters have formed 
usually does not affect this number because
merging and splitting clusters to achieve the new optimal
cluster size imposes a substantial kinetic barrier.
Large bias changes, however, can break up all the clusters and reorganize them into 
clusters with the appropriate number of particles. 
For example, groups of three clusters can merge then split into cluster pairs
with an increase in bias. 
Reducing the applied bias reverses the sequence of transitions,
without obvious hysteresis.

\begin{figure}[t]
  \centering 
  \includegraphics[width=\columnwidth]{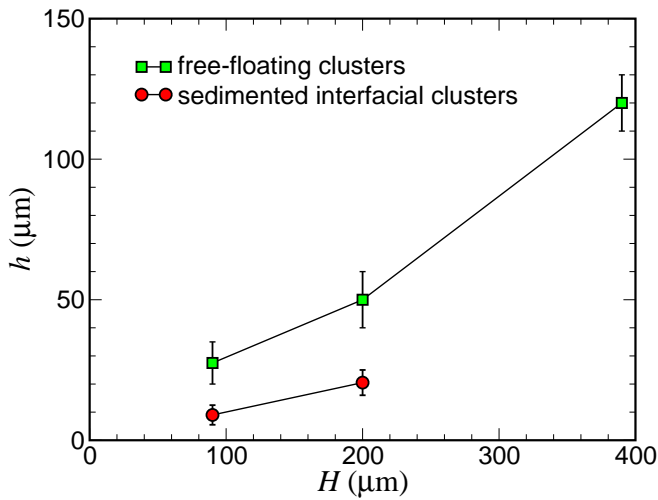}
  \caption{
    Heights of free-floating and sedimented interfacial patterns.
    The error bars represent the variations from run to run.
    In each experiment, all clusters float at the same height,
    with variations well within the error bars.
  }
  \label{fig:height}
\end{figure}

The inter-electrode spacing similarly also not to affect electrokinetic
pattern formation over quite a large range, with similar patterns appearing
in sample cells at $H = 90$, 200, and 390~\micron, all with similar crossover voltages.
Bulk clusters are always observed to float at about $h = H/4$, as shown
in Fig.~\ref{fig:height}.
Interfacial patterns form at a fraction of this height.
Patterns do not form in thinner cells, however.
For example, we observed that particles in a cell $H = 40~\micron$ thick
were pushed directly to the upper electrode and formed no bulk clusters.

\begin{figure}
  \centering 
  \includegraphics[width=\columnwidth]{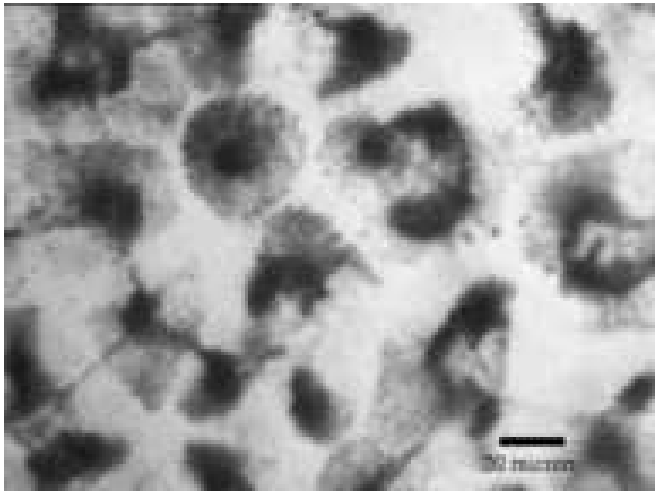}
  \caption{
    3.0~\micron silica spheres in 0.2~mM NaCl solution. $H=200~\micron$. 4~V bias.
    Particles repeated clump into dense clusters then burst out like fireworks.}
  \label{fig:firework}
\end{figure}

Adding salt slows the particles' motions and favors the formation of tumbling clouds instead 
of quasi-stationary clusters.
Patterns are blurred in 0.1~\unit{mM} NaCl solution,
and disappear altogether in 0.5~\unit{mM} solution.
For example, 3.0~\micron silica spheres dispersed in 0.2~\unit{mM} NaCl solution
failed to form any quasi-stationary structures, but rather entered an irregular
cycle of collapsing into concentrated domains and then bursting
like fireworks as shown in Fig.~\ref{fig:firework}.

Unlike the previously reported interfacial crystals, which form for a wide
variety of particle sizes and compositions,
the rich panoply of patterns we have described arises appears only for
particles meeting specific criteria of size and weight.
To clarify this, we compared the electrokinetic pattern formation behavior 
in charge-stabilized dispersions
of colloidal silica spheres, PS spheres and brominated PS spheres
whose densities
are 1.96, 1.05 and 1.6~$\unit{g/cm^3}$, respectively.
The surface charge densities of PS and brominated PS are similar 
and do not depend on the pH.
Both are more highly charged than the silica particles we have discussed so far.
Silica's charge density, moreover, varies with pH, with an isoelectric point
at pH 2.

The behavior of 2.9~\micron diameter silica spheres is very similar to that of
the 3.0~\micron diameter spheres we have discussed to this point. 
These also can form various bulk and low-floating patterns. 
Substantially larger silica spheres never leave the lower electrode at
biases below the threshold for bulk electroconvection.
Substantially smaller, lighter and more diffusive silica spheres create less well
organized patterns.
For example, 1.58~\micron silica particles
usually only form unstable tumbling clouds at comparatively low biases
before being pushed to the upper electrode at higher biases.

From previous studies \cite{Trau97}, it is known that electrokinetic
interactions are weaker for smaller spheres.
This would tend to reduce the forces binding patterns such as colloidal
vortex rings together.
The smaller spheres' more vigorous diffusion also would tend to 
hinder the localized clusters' formation.
It also suggests a possible role for gravity, which can be highlighted
by considering spheres comparable in size to the pattern-forming silica
particles that are composed of different materials.

Unlike silica spheres, the nearly neutrally-buoyant PS spheres fail to form
any bulk patterns at any accessible bias over the entire 
diameter range from 0.3~\micron to 7.0~\micron.
These highly charged spheres should have comparably strong electrokinetic
interactions to their silica counterparts.  Their diffusivity also
should be similar.
Their failure to form bulk patterns demonstrates that the gravity plays
a central role in the pattern formation we have described, a conclusion
consistent with the observation that no bulk patterns form at negative bias.

Dense brominated PS 
spheres 1.6~\micron in diameter
can form tumbling clouds similar to those formed by comparably sized
silica spheres.
Whereas clouds of 1.58~\micron silica spheres tumble vigorously, however,
clouds of brominated PS particles have no detectable internal convection. 
Such quiescent itinerant clouds also appear for dispersions of
3~\micron silica spheres 
but only at much lower concentrations. 
In both cases, each drifting cloud is
about 50~\micron across and its constituent particles are
separated by about 10~\micron.


Stable clusters also do not form at very high or very low particle concentrations.
Dispersions at $\phi\gtrsim 200\%$ form chaotic clouds at low bias and 
jammed clusters at high bias.
The absence of cyclic motion in low-concentration dispersions demonstrates
that the flow patterns we have observed do not arise in the electrolyte alone
under the relevant range of conditions, but rather require the spheres' presence.
The patterns' concentration dependence further highlights the role of cooperative
electrohydrodynamic interactions in establishing the distinct patterns.
Isolated particles' behavior provides insights into these forces
that help to clarify how such patterns form.

\subsection{Single-particle behavior}
\label{sec:single}

To best highlight the role of collective behavior in electrohydrodynamic 
pattern formation, we studied 
the simplest case: a single silica sphere in an electrolysis cell. 
A single charged colloidal sphere of buoyant mass $m$ and charge $q$ 
in a constant electric field not only experiences a downward
gravitational force $mg$ and an electrostatic force $q \vec{E}$
but also more complicated electrohydrodynamic and osmotic forces due to
the electrolyte's response to the field.
When the lower electrode is
negative, the positive screening cloud of the sphere flows downward. 
These flowing ions, including
\hydroxide and \hydronium created by hydrolysis, entrain a flow of water. 
The friction between this electrohydrodynamic flow and the sphere 
drags the sphere downward. 
The sphere's cloud of screening ions is deformed not only by the electric field,
but also by the hydronium flux. This downward
osmotic force and the downward electrohydrodynamic drag are second-order
responses to the electric field, and therefore should be smaller 
than the electrostatic levitating force $q \vec{E}$.

Straightforward considerations allow us to
estimate the bias needed to levitate a particle against gravity.
A single sphere sediments through water at speed
$v_g = mgb = \sigma^2 (\rho - \rho_0) g /(18 \eta)$, where $\rho_0 = 1~\unit{gm/cm^3}$
is the density of water and $\eta = 1~\unit{cP}$ is its viscosity.
For 3~\micron diameter silica spheres, $v_g \approx 5~\unit{\micron/sec}$.
The same particle's electrophoretic velocity saturates at
$v_e \approx e_0 \Sigma \, (V - V_t)/(H \kappa \eta)$,
where $\kappa^{-1}$ is the electrolyte's Debye-H\"uckel screening length,
and $e_0$ is the proton charge \cite{Russel89}.
If we choose the surface charge density $\Sigma_{pH=5}\sim 5000~\unit{\micron^{-2}}$ 
\cite{Behrens01b} and 
$\kappa^{-1}=0.1~\micron$,
these two velocity scales would be equal at about $V - V_t = 0.1~\unit{V}$.
Thus, considering the neglected electrohydrodynamic and osmotic
drags, the threshold for levitation should be at least 0.1~\unit{V}
higher than the water decomposition voltage. This is consistent with our experiments.

\begin{figure}[t]
  \centering 
  \includegraphics[width=\columnwidth]{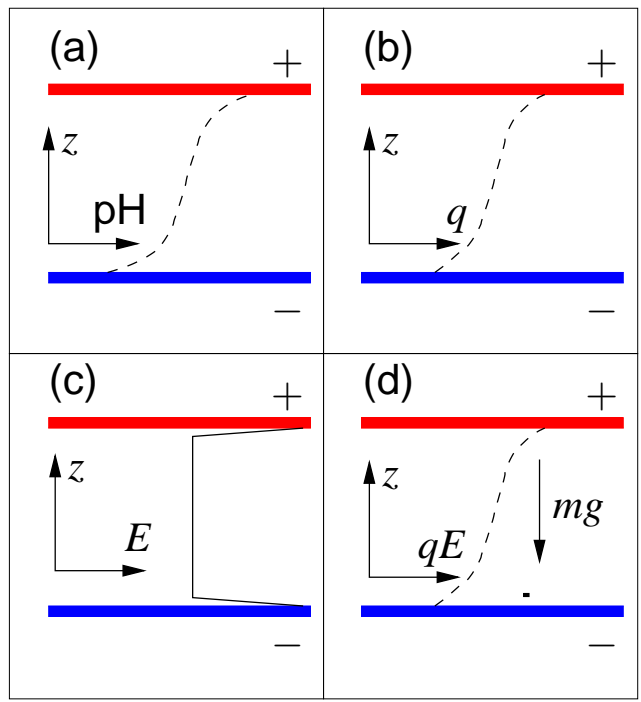}
  \caption{Gravitational and electrostatic forces on a silica sphere.
    (a) Schematically represents the vertical pH gradient established by
    the dissociation of water during hydrolysis.
    (b) The charge $q$ on a sphere tracks variations in pH.
    (c) The electric field $E$ is independent of height outside of the
    Debye-H\"uckel screening layers.
    (d) Even so, the charge-regulating spheres experience an upward force that
    increases with height in the cell.  This force opposes the force due to gravity.
  }
  \label{fig:Eqmg}
\end{figure}
In fact, the pH is not uniform across the cell.
Fig.~\ref{fig:Eqmg} schematically shows how the profiles of 
pH, sphere charge $q$, electric field $E$, and electric force $qE$
vary with height in the cell.
Even though the electric field is uniform in the bulk,
the electrokinetic force levitating a charge-regulating silica 
sphere increases with height.
When this force competes with gravity, therefore, it is possible
to obtain two equilibrium heights, one near the lower electrode
when gravity dominates, and another near the upper electrode
where the electrokinetic force dominates.

The situation is more complicated for two spheres because
hydrodynamic forces between the spheres and bounding surfaces are long-ranged.
Reference~\cite{Squires01} shows the numerically calculated streamlines for one or two 
electrophoretically levitated charged spheres. 
Two particles sinking towards the lower electrode create a back flow which tends to drive them 
apart~\cite{Squires00}. Conversely, two electrically levitated particles will attract each other.
This electrohydrodynamic interaction at least heuristically 
explains why an initially homogeneous colloidal 
fluid breaks up into discrete clusters as it is driven away from the lower electrode.

We observed that the threshold for levitating a single sphere is lower than that
for levitating an entire monolayer.
This most likely reflects the reduced hydrodynamic
drag coefficient for many spheres in a monolayer, as compared to
that for a single sphere \cite{Happel91}.
The increased hydrodynamic attraction generated by the intensified ionic fluxes 
in the more concentrated dispersions would tend to 
draw these sedimented spheres together into clusters. 
This, however, tends to reduce their overall hydrodynamic drag~\cite{Happel91},
causing them to rise less than they otherwise might have.
%
%

\begin{figure}[t]
  \centering 
  \includegraphics[width=\columnwidth]{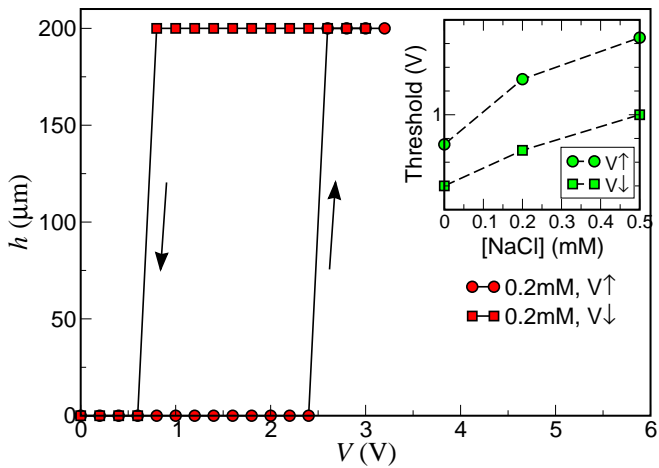}
  \caption{
    Height attained by a single 3~\micron diameter in a $H = 200~\micron$
    cell as a function of applied bias.
    The voltage is increased and then decreased step by step.
    Inset: The threshold voltage required to levitate spheres.}
\label{fig:single}
\end{figure}

To compare our system's behavior with these predictions, we performed experiments on
3~\micron diameter silica spheres from a monolayer with areal density below 
$10^{-4}~\unit{\micron^{-2}}$, \emph{i.e.} fewer than three particles per frame.
Abruptly applying a bias causes these isolated spheres 
to jump off the lower electrode during the transient
in which vertical ionic concentration gradients are established.
Below a threshold voltage,
the particles eventually return to the lower wall.
Above this threshold, they rise to the top wall and remain suspended.
Fig.~\ref{fig:single} shows how the single sphere's height changes with applied voltage 
as the voltage is first increased and then decreased. 
The point at which a single sphere is
electrolevitated to the upper wall is hysteretic: levitated particles only return to the
lower wall when the voltage is lowered substantially. The explanation for this hysteresis 
also can be found in Fig.~\ref{fig:Eqmg}. At the critical bias, $|q \vec{E}(z)| \gtrsim mg$ 
for all $z$ in the bulk. By contrast, a particle can only fall from the upper electrode,
when the field falls below threshold at the upper electrode $|q \vec{E}(H)| \lesssim mg$.
Therefore, the two thresholds are different.
The inset to Fig.~\ref{fig:single} shows this hysteresis more clearly 
at three different salt concentrations. 
Such hysteresis supports the contention that the levitating force acting on
silica spheres increases with the height $h$.

For salt concentration above 1~\unit{mM}, a single particle can never be
forced to the upper electrode.
This is consistent with the observation that particles
move less vigorously and cooperative structures are 
harder to form at high ionic concentrations.

Attractions between particles are detectable only at relatively 
low particle concentrations (\emph{e.g.} $\phi \simeq 10^{-4}$) and 
high applied voltages (\emph{e.g.} 2~\unit{V}). 
In such cases, particles attract each
other as they levitate and so form tumbling clouds. 
The typical inter-particle separation in
the most diffuse clouds is about 10~\micron and decreases with increasing bias.
These field-induced attractions are surprisingly long-ranged, extending to roughly 
50~\micron.
At biases above roughly 5~\unit{V}, even isolated spheres
travel in circular orbits whose diameters range
from  50 to 150~\micron in the vertical plane. 
This motion is due to electroconvection in the underlying electrolyte.
It almost certainly plays no role in forming the highly organized
and substantially smaller-scale patterns discussed above.

Generally speaking, single particles are not stably levitated 
into the bulk of the electrolyte,
and they certainly do not trace out the complex trajectories
characterizing microscopic patterns such as those in 
Figs.~\ref{fig:vortexring} - \ref{fig:firework}.
These observations help to confirm that such patterns
do not arise from electroconvection in the electrolyte alone.
We are left to conclude that
the spheres play an active role in dynamical pattern selection.

\section{Discussion}

We have demonstrated that many distinctive patterns reproducibly emerge from the interplay 
of gravity and electrohydrodynamic forces in charge-stabilized colloid. 
At biases above
the threshold for hydrolysis but below the onset of electroconvection, only fluxes of
\hydronium and \hydroxide ions flow through the system. 
Electroneutrality favors overlap of these fluxes; however, ``friction'' 
between opposite flows is minimized if these flows are separated.
The optimal balance between these effects separates the fluxes by a characteristic
lengthscale comparable to the Debye-H\"uckel screening length, or roughly 0.1~\micron.
This is much smaller than the extent of the experimentally observed
clusters whose diameters extend to roughly 100~\micron.
Electroneutrality suppresses the formation of such large-scale convective structures
in the absence of the charged spheres.
In other words, are not simply passive tracers for a pre-existing flow in the
electrolyte, but rather they and their charges are essential for establishing
the convective flows at the heart of the observed patterns.

There are three general features about the dynamics of all clusters.
(1) They have the same circulating direction as shown in Fig.~\ref{fig:schematic}.
(2) The circulation speed increases with the bias as clusters tend to become more compact.
(3) The number of particles in each cluster increases with the bias. The bias-induced attraction of the second feature is crucial to understand the
clusters' formation. The breathing mode in Figs.~\ref{fig:oscillate} and \ref{fig:firework}
may result from competition between this as-yet unexplained attraction and a combination of
screened-Coulomb repulsions and induced dipole repulsions among spheres.

While the sequence of dynamical transitions from sedimented layer
to tumbling clouds to highly organized clusters is reproducible, 
the crossover potentials depend sensitively on the changing
electrochemical properties of the electrodes, and therefore
can vary by as much as $\pm 0.4~\unit{V}$ from run to run.

\section{Summary}

We have reported a large family of previously unrecognized 
self-organized colloidal patterns that form in constant vertical
electric fields at biases just above the decomposition voltage of 
water. Our simple electrolysis cell gives rise to a complicated coupled 
system of hydrodynamic flows and ionic fluxes
that are shepherded by the very colloidal particles they transport.
We have discussed the electrohydrodynamic forces on a single particle
in this system, and suggested how many-body coupling might give
rise to the various dynamical structures we have observed.
At low voltages, colloidal spheres in a specific range of charges, 
densities and mobilities are found to
form various microscopic quasi-steady-state clusters through the competition of gravity and 
electrohydrodynamic levitating forces. 
These patterns falls into two categories: clusters levitated into the bulk 
and clusters localized near the lower electrode. 
Their structures and evolutions are different, 
but all share some general features, such as persistent circulation
and coarsening with increasing voltage. 
The simple behavior of a single particle in the field 
demonstrates that these microscopic clusters are formed cooperatively. 
Patterns' dependence on such control parameters as the composition 
and number density of particles, cell thickness $H$,
salt concentration and applied voltage range have been discussed.

This work was supported by the Donors of the Petroleum Research Fund
of the American Chemical Society.

%

\begin{thebibliography}{10}

\bibitem{Russel89}
W.~B. Russel, D.~A. Saville, and W.~R. Schowalter, {\em Colloidal Dispersions},
  {\em Cambridge Monographs on Mechanics and Applied Mathematics} (Cambridge
  University Press, Cambridge, 1989).

\bibitem{Muth27}
E. Muth, Kolloid Zeitschrift {\bf 41},  97  (1927).

\bibitem{Winslow49}
W.~M. Winslow, J. Appl. Phys. {\bf 20},  1137  (1949).

\bibitem{Jennings90}
B.~R. Jennings and M. Stankiewicz, Proc. Roy. Soc. London {\bf 427},  321
  (1990).

\bibitem{Hu94}
Y. Hu, J.~L. Glass, A.~E. Griffith, and S. Fraden, J. Chem. Phys. {\bf 100},
  4674  (1994).

\bibitem{Isambert97}
H. Isambert, A. Ajdari, J.~L. Viovy, and J. Prost, Phys. Rev. Lett. {\bf 78},
  971  (1997).

\bibitem{Isambert97a}
H. Isambert, A. Ajdari, J.-L. Viovy, and J. Prost, Phys. Rev. E {\bf 56},  5688
   (1997).

\bibitem{Larsen96}
A.~E. Larsen, Ph.d. thesis, The University of Chicago, 1996.

\bibitem{Larsen96a}
A.~E. Larsen and D.~G. Grier, Phys. Rev. Lett. {\bf 76},  3862  (1996).

\bibitem{Larsen97}
A.~E. Larsen and D.~G. Grier, Nature {\bf 385},  230  (1997).

\bibitem{Richetti84}
P. Richetti, J. Prost, and P. Barois, J. Physique Lett. {\bf 45},  L1137
  (1984).

\bibitem{Wei93}
Q.~H. Wei, X.~H. Liu, C.~H. Zhou, and N.~B. Ming, Phys. Rev. E {\bf 48},  2786
  (1993).

\bibitem{Lei95}
X.~Y. Lei {\it et~al.}, Phys. Rev. E {\bf 52},  5161  (1995).

\bibitem{Wei95}
Q.~H. Wei, X.~Y. Lei, C.~H. Zhou, and N.~B. Ming, Phys. Rev. E {\bf 51},  1586
  (1995).

\bibitem{Ristenpart03}
W.~D. Ristenpart, I.~A. Aksay, and D.~A. Saville, Phys. Rev. Lett. {\bf 90},
  128303  (2003).

\bibitem{Yethiraj03}
A. Yethiraj and A. Blaaderen, Nature {\bf 421},  513  (2003).

\bibitem{Trau96}
M. Trau, D.~A. Saville, and I.~A. Aksay, Science {\bf 272},  706  (1996).

\bibitem{Trau97}
M. Trau, D.~A. Saville, and I.~A. Aksay, Langmuir {\bf 13},  6375  (1997).

\bibitem{Bohmer96}
M. B\"{o}hmer, Langmuir {\bf 12},  5747  (1996).

\bibitem{Solomentsev97}
Y. Solomentsev, M. B\"{o}hmer, and J.~L. Anderson, Langmuir {\bf 13},  6058
  (1997).

\bibitem{Duan01}
X.~D. Duan and W.~L. Luo, Int. J. Mod. Phys. B {\bf 15},  837  (2001).

\bibitem{Han03a}
Y. Han and G.~G. Grier, Nature {\bf 424},  267  (2003).

\bibitem{Behrens01b}
S.~H. Behrens and D.~G. Grier, J. Chem. Phys. {\bf 115},  6716  (2001).

\bibitem{Bockris70}
J.~O. Bockris and A.~K.~N. Reddy, {\em Modern Electrochemistry}
  (Plenum/Rosetta, New York, 1970).

\bibitem{Kortum65}
G. Kort\"um, {\em Treatise on Electrochemistry} (Elsevier Pub. Co., Amsterdam,
  1965).

\bibitem{Nadal02}
F. Nadal {\it et~al.}, Phys. Rev. E {\bf 65},  061409  (2002).

\bibitem{Shariff92}
K. Shariff and A. Leonard, Annu. Rev. Fluid Mech. {\bf 24},  235  (1992).

\bibitem{Squires01}
T.~M. Squires, J. Fluid Mech. {\bf 443},  403  (2001).

\bibitem{Happel91}
J. Happel and H. Brenner, {\em Low Reynolds Number Hydrodynamics} (Kluwer,
  Dordrecht, 1991).

\bibitem{Squires00}
T.~M. Squires and M.~P. Brenner, Phys. Rev. Lett. {\bf 85},  4976  (2000).

\end{thebibliography}

\end{document}